\documentstyle[preprint,pra,aps,psfig]{revtex}

\begin{document}
\draft

\title{Ortho-Para Conversion in CH$_3$F.\\
 Self--Consistent Theoretical Model\thanks{Presented at
 the VI International Symposium on Magnetic Field and Spin Effects in
 Chemistry and Related Phenomena, Emmetten, Switzerland, August 21-26, 1999.}}

\author{ Pavel L. Chapovsky\thanks{E-mail: chapovsky@iae.nsk.su}} 
\address{\sl Institute of Automation and
Electrometry, 
The  Russian Academy of Sciences,\\ 630090 Novosibirsk, Russia;
and Laboratoire de Physique des Lasers, 
Universit\'e des Sciences et Technologies de Lille, 
F-59655 Villeneuve d'Ascq Cedex, France}
\date{\today}
\maketitle

\begin{abstract}
A complete theoretical model of the nuclear spin conversion in 
$^{13}$CH$_3$F induced by intramolecular ortho-para state mixing is proposed. 
The model contains parameters determined from the 
level-crossing spectra of the $^{13}$CH$_3$F spin conversion. This set 
of  parameters includes the ortho-para decoherence rate, the magnitude 
of the hyperfine spin-spin interaction between the
molecular nuclei and the energy gap between the mixed ortho and para states. 
These parameters are found to be in a good agreement with their 
theoretical estimates.
\end{abstract}


\section{Introduction}

The study of nuclear spin isomers of molecules was started
by the discovery of the ortho and para 
hydrogen in the late 1920's \cite{Farkas35}.  
It became clear already in that time that many other symmetrical molecules should
have nuclear spin isomers too. Nevertheless, 
their investigation has been postponed by almost 
60 years. The reason for this delay was severe difficulties
in the enrichment of spin isomers. The situation is improving 
now (see the review in Ref.~\cite{Chap99ARPC}) but yet we are at 
the very early stage of this  research: in addition to the well-known 
spin isomers of H$_2$ only a few molecules have been investigated so far. 
Among them, the CH$_3$F nuclear spin isomers occupy a special place being the 
most studied and understood.

The conversion of CH$_3$F nuclear spin isomers has been  
explained \cite{Chap90JETP,Chap91PRA}
in the framework of {\it quantum relaxation} \cite{Chap96PA}. 
which is based on the intramolecular ortho-para state mixing 
and on the interruption of this mixing by collisions. This mechanism 
of spin conversion has a few striking features. 
The nuclear spin states  of  CH$_3$F appeared to be extremely stable
surviving $10^9 - 10^{10}$ collisions. 
Each of the collision changes the energy of the molecule by 
$10 - 100$~cm$^{-1}$ and shuffles the molecular rotational state substantially. 
Nevertheless, the model predicts that  the
spin conversion is governed by tiny intramolecular interactions 
having the energy $\sim10^{-6}$~cm$^{-1}$.

Under these circumstances, the validity of the proposed theoretical 
model should be checked with great care. 
This is especially important because the 
CH$_3$F case gives us the first evidence of the new mechanism
behind the nuclear spin conversion in molecules. Hydrogen spin conversion,
which is the only other comprehensively studied case, is 
due to the completely different process based on direct  
collisional transitions between ortho and para states of H$_2$.

Presently there is substantial amount of the experimental data on
CH$_3$F isomer conversion (see \cite{Chap99ARPC} and references therein).
Theory and experiment on the CH$_3$F isomer conversion were compared 
in a number of papers but these comparisons were never aimed to 
determine a complete set of parameters
necessary for {\it a quantitative} description of the process.  
The purpose of the present paper is to construct such self-consistent 
theoretical model of the CH$_3$F isomer conversion.

\section{Quantum relaxation}

\begin{figure}[htb]
\centerline{\psfig
{figure=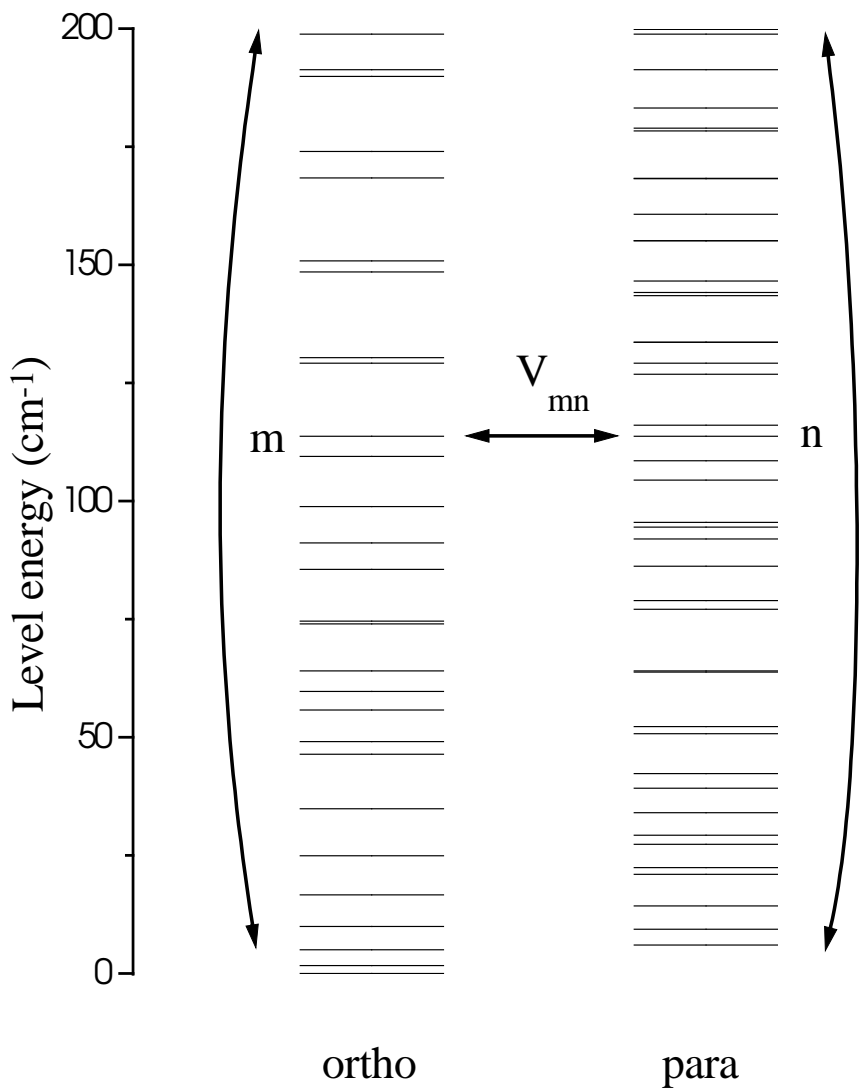,height=18cm}}
\vspace{0.5cm}
\caption{\sl \label{fig1} Ortho and para states of $^{13}$CH$_3$F
in the ground vibrational state. 
The level energies were calculated using the molecular parameters 
from [17]. The (11,1)--(9,3)
pair of states is shown to be mixed by the intramolecular perturbation. 
The bent lines indicate the rotational relaxation induced
by gas collisions. These collisions do not provide direct ortho-para
transitions.} 
\end{figure}

The CH$_3$F molecule
is a symmetric top having the C$_{3v}$ symmetry. 
The total spin of the three
hydrogen nuclei in the molecule can be equal to $I=3/2$ (ortho isomers), 
or $I=1/2$ (para isomers). The values of the molecular angular momentum
projection on the molecular symmetry axis ($K$) are specific for these 
spin isomers. Only $K$ divisible by 3 are allowed for the ortho isomers. 
All other $K$ are allowed for the para isomers. Consequently, the 
rotational states of CH$_3$F form two subspaces
which are shown in Fig.~\ref{fig1} for the particular
case of the $^{13}$CH$_3$F molecules.

Let us briefly recall the physical picture of the CH$_3$F
spin conversion by quantum relaxation. 
Suppose that a test molecule was placed initially in the 
ortho subspace of the molecular states (Fig.~\ref{fig1}). Due to collisions 
in the bulk the test molecule will undergo fast rotational relaxation 
{\it inside} the ortho subspace. This running up and down
along the ortho ladder proceeds until the molecule reaches  the
ortho state $m$ which is mixed with the para state $n$ by the 
{\it intramolecular} perturbation $\hat V$. Then, during the free flight
just after this collision, the perturbation $\hat V$ mixes the para state $n$
with the ortho state $m$. Consequently, the next collision is able
to move the molecule to  other para states and thus to localize
it inside the para subspace. Such  mechanism of spin isomer conversion was 
proposed in the theoretical paper \cite{Curl67JCP}.    

The quantum relaxation of spin isomers can be quantitatively
described in the framework of the kinetic equation for 
density matrix \cite{Chap90JETP}. Let us consider first a free molecule
which is not subjected to an external field. One needs to split the
molecular Hamiltonian into two parts
\begin{equation}
              \hat H = \hat H_0 + \hbar\hat V,
\label{H}              
\end{equation}
where the main part of the Hamiltonian, $\hat H_0$, has pure ortho 
and para states as the eigenstates; the perturbation $\hat V$
mixes the ortho and para states. In the
first order perturbation theory the nuclear spin conversion rate, 
$\gamma$, is given by
\begin{equation}
        \gamma = \sum_{\alpha'\in p, \alpha\in o}
        \frac{2 \Gamma_{\alpha'\alpha}|V_{\alpha'\alpha}|^2}
        {\Gamma^2_{\alpha'\alpha} + \omega^2_{\alpha'\alpha}}
        \left(W_p(\alpha') + W_o(\alpha)\right),
\label{gamma}        
\end{equation}
where $\Gamma_{\alpha'\alpha}$ is the decay rate of the 
off-diagonal density matrix element $\rho_{\alpha'\alpha}\ 
(\alpha'\in para;\ \ \alpha\in ortho)$; $\hbar\omega_{\alpha'\alpha}$ is 
the energy gap between the 
states $\alpha'$ and $\alpha$; $W_p(\alpha')$ and $W_o(\alpha)$ are 
the Boltzmann factors of the corresponding states.
The parameters $\Gamma_{\alpha'\alpha}$, $V_{\alpha'\alpha}$, and
$\omega_{\alpha'\alpha}$ are crucial for the quantitative theoretical
description of the $^{13}$CH$_3$F spin isomer conversion.

All previous comparisons between 
the experiment on the CH$_3$F spin conversion and  
the theory were performed using {\it ``total''} rates of conversion 
which summarize all contributions to the rate from
many ortho-para level pairs. The ``total''
rate is just a single number and obviously cannot provide
unambiguous determination of all parameters 
which are present in the expression (\ref{gamma}). One may combine the
experimental data on  ``total'' rates with theoretical calculations
of some parameters but it is not easy. In this case one has to perform extensive 
calculations of  the intramolecular ortho-para state mixing. 
Even more difficult is to calculate 
the decoherence rates $\Gamma_{\alpha'\alpha}$. Consequently, 
development of the self-consistent model 
of  the nuclear spin conversion in which all parameters are
unambiguously determined should be based on a different approach.  

\section{Level--crossing resonances}

Theoretical model of spin conversion predicts strong dependence of the 
conversion rate, $\gamma$, on the level spacing $\omega_{\alpha'\alpha}$ 
(see Eq.~(\ref{gamma})). This can be used to single out
the contribution to the conversion from each level pair
which should substantially simplify the quantitative comparison
between theory and experiment. It was proposed in \cite{Nagels95CPL}  
and performed in \cite{Nagels96PRL} to use the Stark effect for 
crossing the ortho and para states of CH$_3$F. These crossings 
result in sharp increase of the conversion rate $\gamma$ giving
the conversion spectra if electric field is varied. The experimental data
\cite{Nagels96PRL} are presented in Fig.~\ref{fig2}.  
It is evident that such
spectrum contains much more information than the "total" 
conversion rate which is just a single number. 

Comparison of the conversion spectrum in Fig.~\ref{fig2} 
with the theory needs a modification of the model in order to 
incorporate the Stark effect. 
Homogeneous electric field lifts partially 
the degeneracy of the $\alpha$-states
of CH$_3$F (see Appendix). The new states, $\mu$-basis, can 
be found in a standard way \cite{Landau77}:
\begin{equation}
     |\mu>\ \equiv\ |\beta,\xi>|\sigma_F>|\sigma_C>;\ \ \xi=0,1.
\label{m}
\end{equation}
Because electric field in the experiment is relatively small, it is
sufficient to consider only diagonal matrix elements of the Stark
perturbation over angular momentum $J$ when calculating the $\mu$-states.
Energy of  the $\mu$-states are given by the expression
\begin{equation}
     E(\mu) = E_{free}(J,K) + (-1)^\xi \frac{K|M|}{J(J+1)}|d{\cal E}|,
\label{E}
\end{equation}
where $E_{free}(J,K)$ is the energy of free molecule; $d$ is the molecular 
permanent electric dipole moment; ${\cal E}$ is the electric field strength. 
The new states are still degenerate with respect to the 
spin projections $\sigma$, $\sigma_F$, and $\sigma_C$, and to the
sign of $M$. An account of the Stark effect in the spin conversion model 
is straightforward. 
Eq.~(\ref{gamma}) should be rewritten in the $\mu$-basis with the 
level energies determined by the Eq.~(\ref{E}).

\section{Fitting of the experimental data}

Nuclear spin conversion in $^{13}$CH$_3$F at zero electric field
is governed almost completely by mixing of only two level
pairs ($J'$=11,~$K'$=1)--($J$=9,~$K$=3) and (21,1)--(20,3)
\cite{Chap91PRA,Chap93CPL}.
The spectrum presented in Fig.~\ref{fig2} is produced by  crossings of the 
$M$-sublevels of the para (11,1) and ortho (9,3) states. This pair 
of states is mixed by the spin-spin interaction between the molecular nuclei
\cite{Chap91PRA}. There is no contribution to the mixing of this 
level pair from the spin-rotation interaction because of the selection rule
for spin-rotation interaction $|\Delta J|\leq 1$ \cite{Guskov95JETP}. 
This is fortunate because the spin-spin 
interaction can be calculated rather accurately. Contrary to that, the spin-rotation  
interaction in CH$_3$F is known only approximately.
For more details on spin-rotation contribution to the CH$_3$F spin conversion
see Refs.~\cite{Guskov95JETP,Chap97,Bahloul98JPB,Ilisca98PRA,Guskov99JPB}

The second pair of ortho-para states (21,1)--(20,3), which is also
important for the spin conversion in $^{13}$CH$_3$F at zero electric field,  
is mixed by both the spin-spin and 
spin-rotation interactions. The magnitude of the latter is presently 
unknown. Nevertheless, it does not complicate the fitting procedure because 
in the vicinity of the (11,1)--(9,3) resonances presented in Fig.~\ref{fig2},
the (21,1)--(20,3) pair gives very small and almost constant contribution. 

Let us find out now an analytical expression for modelling 
the experimental data. We start by analyzing
the contribution to the conversion rate produced by 
the level pair (11,1)--(9,3) which will be  denoted  as $\gamma_a({\cal E})$.
This contribution
can be obtained using the results of Refs.~\cite{Chap91PRA,Nagels95CPL}:
\begin{eqnarray}
     \gamma_a({\cal E}) &=& \sum_{M'\in p;\ M\in o}
      \frac{2\Gamma |V_{M'M}|^2}{\Gamma^2 + \omega^2_{M'M}({\cal E})}
                   \left(W_p(\mu') + W_o(\mu)\right);  \nonumber \\
     |V_{M'M}|^2 &=& (2J+1)(2J'+1)
                          \left(\begin{array}{rcc} 
                                J' & J & 2 \\
                               -K' & K & K'-K\\
                             \end{array}\right)^2
                       \left(\begin{array}{rcc} 
                               J' & J & 2 \\
                              -M' & M & M'-M\\
                             \end{array}\right)^2
                                {\cal T}^2.
\label{gE}
\end{eqnarray}
Here $V_{M'M}\equiv <\mu'|V|\mu>$ are the matrix elements of the
perturbation $\hat V$ in which only $M$-indexes were 
shown explicitly;
(:::) stands for the 3j-symbol; ${\cal T}$ is the magnitude of the 
spin-spin interaction. Note, that the selection
rules for the ortho-para state mixing by the spin-spin interaction
result from Eq.~(\ref{gE}): $|\Delta K|; |\Delta J|; |\Delta M|\leq2$. 
In the fitting procedure ${\cal T}$
will be considered as an adjustable parameter. 
In Eq.~(\ref{gE}) we have assumed all 
$\Gamma_{M'M}$ being equal: $\Gamma_{M'M}\equiv \Gamma$. 
This property of $\Gamma$ is the consequence of the spherical
symmetry of the media. The decoherence decay rate $\Gamma$ is an another
unknown parameter which needs to be determined. 

The spacing between the $M'$ and $M$ states in an electric filed 
follows directly from the Eq.~(\ref{E}) 
\begin{equation}
     \omega_{M'M}({\cal E}) = \omega_0 + 
     \left(\frac{K'|M'|}{J'(J'+1)} - \frac{K|M|}{J(J+1)}\right)
     |d{\cal E}|,
\label{om}
\end{equation}
where $\omega_0$ is the gap between the states ($J'$,$K'$) and ($J$,$K$) 
at zero electric field. We have considered in Eq.~(\ref{om})
only pairs of states which have $\xi'=\xi$. They are the only pairs
which contribute to the spectrum in the electric field range
of Fig.~\ref{fig2}. The level spacing $\omega_0$ will be 
considered  as an adjustable parameter in 
the fitting. The dipole moment of $^{13}$CH$_3$F in the 
ground state, which is necessary for the calculation of $\omega_{M'M}({\cal E})$, was 
determined very accurately from the laser Stark spectroscopy of 
$^{13}$CH$_3$F and was found equal $d=1.8579~\pm~0.0006$~D 
\cite{Freund74JMS}.

At zero electric field the  level pair (21,1)--(20,3) contributes 
nearly 30\% to the total conversion rate \cite{Chap93CPL}. At  
electric fields, where 
$\gamma_a({\cal E})$ has peaks, 
this contribution is on the order of 10$^{-2}$  in comparison with  
$\gamma_a({\cal E})$. The first crossing of the pair (21,1)-(20,3) occurs at 
$\simeq 4000$~V/cm thus having its peaks far away 
from the electric field range of Fig.~\ref{fig2}. In the electric field range of
Fig.~\ref{fig2}  (1--1200~V/cm) the contribution from the 
pair (21,1)-(20,3) is changing 
by 10\% only. Consequently, in the fitting procedure the (21,1)-(20,3) 
contribution is assumed to be constant.
This quantity will be denoted as $\gamma_b$.

To summarize, the function which will be used to model the experimental 
data is
\begin{equation}
     \gamma({\cal E}) = \gamma_a({\cal E}) + \gamma_b.
\label{gEg}
\end{equation}  
This function contains adjustable parameters  ${\cal T}$, 
$\Gamma$, $\omega_0$, and $\gamma_b$. 

The result of the least-square fit is shown in Fig.~\ref{fig2} by solid line.
The error of the individual experimental points in Fig.~\ref{fig2} was 
estimated as 7\%. The values of the parameters are given in the
Table~1, where one standard deviation of statistical error is indicated.

Electric field in the Stark cell was determined in experiment \cite{Nagels96PRL}
by measuring the voltage applied to the electrodes and assuming the
distance between them equal to 4.18 mm, which is the spacer thickness.
It was found out after the experiment \cite{Nagels96PRL} was performed that
the thickness of the glue used to attach the Stark electrodes was not
negligible. The updated spacing between the electrodes in the Stark cell is
$l=4.22\pm 0.02$~mm. Such correction of the spacing gives 1\% 
systematic decrease of the experimental electric field values given in 
\cite{Nagels96PRL}. This shift is taken into account in Fig.~\ref{fig2}.

\section{Theoretical estimation of the parameters}

Let us compare the parameters obtained in the previous section
with their theoretical estimates. We start from the analysis of the 
level spacing $\omega_0$. The best sets of the ground state 
molecular parameters of $^{13}$CH$_3$F are given  in Ref.~\cite{Papousek94JMS}. 
The spacing between the levels (11,1) and (9,3) is presented in 
the Table~1 where the set having most accurate molecular 
parameter $A_0$ was used. The theoretical value appears to be close 
to the experimental one obtained from the spin conversion spectra. 
The difference between them is
\begin{equation}
     \omega_0(exp)-\omega_0(theor) = 1.0 \pm 0.3\ {\text{MHz}},
\label{oo}
\end{equation}
which is less than 1\% in comparison with $\omega_0$ itself.

Next, we calculate the parameter ${\cal T}$ which 
characterizes the spin-spin mixing of the level pair
(11,1)--(9,3) in $^{13}$CH$_3$F. The spin-spin interaction 
between the two magnetic dipoles ${\bf m}_1$ and
${\bf m}_2$ separated by the distance ${\bf r}$ has the form \cite{Landau77}:
\begin{eqnarray}
     \hbar \hat V_{12} &\ =\ & 
     P_{12}\hat{\bf I}^{(1)}\hat{\bf I}^{(2)}
     \ ^\bullet_\bullet\ {\text{\bf T}}^{(12)}\ , \nonumber \\
     T_{ij}^{(12)}     &\ =\ &
     \delta _{ij}-3n_in_j\ ;\phantom{TT}P_{12}=
     m_1m_2/r^3I^{(1)}I^{(2)}\ ,
\label{ss12}     
\end{eqnarray}
where  $\hat{\bf I}^{(1)}$ and $\hat{\bf I}^{(2)}$ are the spin 
operators of the particles 1 and 2, respectively; {\bf n} is the unit 
vector directed along
{\bf r}; $i$ and $j$ are the Cartesian indexes.

For the spin-spin mixing of the ortho and para states in $^{13}$CH$_3$F
one has to take into account the interaction between the three hydrogen nuclei
($\hat V_{HH}$), between the three hydrogen  and fluorine 
nuclei ($\hat V_{HF}$), and between the three hydrogen and  carbon nuclei
($\hat H_{HC}$). Thus the total spin-spin interaction 
responsible for the mixing in $^{13}$CH$_3$F is
\begin{equation}
              \hat V_{SS} = \hat V_{HH} + \hat V_{HF} + \hat V_{HC}.
\label{VSS}
\end{equation}
The complete expressions for all components of $\hat V_{SS}$ 
can be written by using Eq.~(\ref{ss12}) 
for the spin-spin interaction between two particles. 
For example, for $\hat V_{HF}$ one has
\begin{equation}
               \hat V_{HF}\ =\ P_{HF}\sum_{n}\hat{\bf I}^{(n)}
          \hat{\bf I}^{F}\ {}^\bullet_\bullet\ {\bf T}^{nF}\ ;\ \ n=1,2,3\ .
\label{HF}          
\end{equation}
Here $P_{HF}$ is the scaling factor analogous to $P_{12}$ in Eq.~(\ref{ss12});
$n$ refers to the hydrogen nuclei in the molecule. 

${\cal T}$ can be calculated in a way similar to that used previously 
\cite{Chap91PRA}. It gives
\begin{equation}
     |{\cal T}|^2 = 3|P_{HH}{\cal T}^{(12)}_{2,2}|^2 +
                      2|P_{HF}{\cal T}^{1F}_{2,2}|^2 +
                         2|P_{HC}{\cal T}^{1C}_{2,2}|^2.
\end{equation}
Here ${\cal T}^{1q}_{2,2}$ is the spherical component of the second 
rank tensor {\bf T}$^{1q}$ calculated in the molecular system of coordinates. 
The superscripts $1q$ indicate the interacting particles: 1 refers to the
hydrogen nucleus H$^{(1)}$ and $q$ refers to the
nucleus of H$^{(2)}$, or F, or C. 

The calculation of ${\cal T}$ needs the knowledge of the molecular structure.
We used the ground state structure of $^{13}$CH$_3$F determined in
\cite{Egawa87JMStr,Egawa98}: $r_{CF} = 1.390(1)$~\AA, $r_{CH} = 1.098(1)$~\AA,
and $\beta(F-C-H)=108.7^o(2)$. The numbers in parentheses represent the
error bars in units of the last digit. 
By using these parameters one can obtain the value of 
${\cal T}$ which is given in the Table~1. The difference between the 
experimental and theoretical values of ${\cal T}$ is equal to
\begin{equation}
     {\cal T}_{exp} - {\cal T}_{theor} = -5.1 \pm 0.5\  {\text{kHz}}.
\label{T}
\end{equation}

\section{Discussion}

Small difference between the 
experimental and theoretical values of $\omega_0$ unambigiously
confirms that the mixed ortho-para level pair (9,3)--(11,1)
was determined correctly.
From the spectroscopical data \cite{Papousek94JMS} one can conclude
that there are no other ortho-para level pairs which 
can mimic the level spacing $\omega_0=130.99$~MHz.
It is also true if one 
takes into account even all ortho-para level pairs ignoring the
restrictions imposed by the selection rules for the
ortho-para state  mixing.

The difference between experimental and theoretical values of
the level spacing at zero electric field,
$\omega_0$, is only $1.0\pm0.3$~MHz.
The main error in theoretical value of $\omega_0$ is caused
by the error in the molecular parameter $A_0$. It gives nearly
half of the error indicated in the Table~1. On the other hand,
the $J$ and $K$ dependences of the molecular electric dipole
moment are too small \cite{Freund74JMS} to affect our determination of the
theoretical value of $\omega_0$. It is possible that the
experimental value of $\omega_0$ is affected by the pressure shift,
which magnitude we presently do not know. Further investigations
can precise the frequency gap between the
states (9,3) and (11,1).

The difference between the experimental and theoretical values
of ${\cal T}$ is rather small ($\simeq$7\%) but well outside the
statistical error. This difference
may originate from our method of calculating ${\cal T_{theor}}$ in  
which we used the molecular structure (bond lengths
and angles) averaged over ground state molecular vibration. More correct
procedure would be to average an exact expression
for ${\cal T}$ over molecular vibration. This requires rather 
extensive calculations. 

There are a few contributions to the systematic error of 
value of ${\cal T_{exp}}$. The response time of the
setup used to measure the concentration of ortho molecules ($\simeq1$~sec)
was not taken into account in the processing of the experimental data. 
This gives $\simeq$2\% systematic decrease
in the value of ${\cal T_{exp}}$. 
Another few percent of the systematic error may appear due to the
procedure employed in \cite{Nagels96PRL} to find out the conversion rate
inside the Stark cell. This procedure relies on 
the ratio of the Stark cell volume to the volume
outside the electric field. Taking  these
circumstances into account we can estimate that up to $\simeq$10\% difference between
the experimental and theoretical values of ${\cal T}$ can be explained
by the systematic errors.  Despite this difference,
it is rather safe to conclude that our analysis has proven that
the levels (9,3) and (11,1)
are indeed mixed by the spin-spin interaction between the
molecular nuclei. It is impressive that the level-crossing spectrum
in the $^{13}$CH$_3$F isomer conversion has
allowed to  measure the hyperfine spin-spin coupling with
the statistical error of 0.5 kHz only.

Comparison between the measured spectrum and the model supports
our choice for the $\Gamma_{M'M}$ being independent on $M$ and $M'$.
Independence of this parameter on $M$ is the direct
consequence of the spatial isotropy of the media. The 
independence on $M'-M$ is more intricate. This will be discussed 
in more detail elsewhere.

The value of $\Gamma$ obtained from the fitting procedure, 
$\Gamma=(1.9 \pm 0.1)\cdot10^8$~s$^{-1}$/Torr, appeared 
to be close to the level population decay rate 
$1.0\cdot10^8$~s$^{-1}$/Torr measured in Ref.~\cite{Jetter73JCP} for the state 
($J$=5,~$K$=3) of $^{13}$CH$_3$F. The factor 2 difference is not 
surprising. $\Gamma$ refers to the decay rate of 
the off-diagonal density matrix element $\rho_{\mu'\mu}$
between the states (11,1) and (9,3) which should be
different from the population decay rate. In addition, the 
rotational quantum numbers in these two cases are different too.  

Column designated as $\gamma(0)$ in the Table~1 gives the rates 
at zero electric filed. The ``theoretical value'' is the magnitude of
$\gamma(0)$ given by the solid line in Fig.~\ref{fig2}.
The theoretical value coincides well with
the experimental one from Ref.~\cite{Nagels96PRA54}.
Finally we would like to mention that our analysis of the spin 
conversion spectrum has allowed to disentangle for the first time 
the contributions to the conversion rate which arise from the 
mixing of the two level pairs: (9,3)--(11,1) and (20,3)--(21,1). 

\section{Conclusions}

We have performed the first quantitative comparison of the 
level-crossing spectrum of the
nuclear spin conversion in $^{13}$CH$_3$F with the theoretical model.
This approach has allowed to single out the contribution to
the spin conversion caused by the mixing of one particular 
pair of the ortho-para rotational states of the molecule and 
confirmed unambiguously that the mechanism of the intramolecular 
state mixing is the spin-spin interaction between the molecular nuclei.

All important
parameters of the theoretical model which describe the
nuclear spin conversion in $^{13}$CH$_3$F due to the spin-spin
mixing of the ortho-para level pair (9,3)--(11,1) are determined 
quantitatively. These parameters are the decoherence rate, $\Gamma$,
the spin-spin mixing strength, ${\cal T}$, the level spacing, $\omega_0$,
and the contributions to the conversion rate from the two
level pairs separately (9,3)-(11,1) and (20,3)-(21,1).
While the decoherence rate $\Gamma$ is difficult to estimate
on the basis of
independent information, the experimental values for the spin-spin
mixing, ${\cal T}$, and the level spacing, $\omega_0$, are found to be close
to their theoretical values. These results 
prove that the nuclear spin conversion in the $^{13}$CH$_3$F
molecules is indeed governed by the quantum relaxation. 

\section*{Acknowledgments}

This work was made possible by financial support from the the 
Russian Foundation 
for Basic Research (RFBR), grant No. 98--03--33124a, and the 
R\'egion Nord Pas de Calais, France.

\section{Appendix}

The CH$_3$F quantum states in the ground electronic and vibration 
state can be classified as follows \cite{Townes55,Bunker79,Cosleou99EPJD}. 
CH$_3$F is a rigid symmetric top but it is more transparant to 
take molecular inversion into account and classify the states in 
D$_{3h}$ symmetry group. First, one has to introduce 
an additional (molecular) system of coordinates which has 
the orientation defined by 
the numbered hydrogen nuclei and $z$--axis directed along the molecular 
symmetry axis.

Next, one introduces the states 
\begin{equation}
          |\beta>\ \equiv\ |J,K,M>|I,\sigma,K>;\ \ K\geq0,
\label{b}
\end{equation}
which are invariant under cyclic permutation of the three hydrogen nuclei:
$P_{123}|\beta>=|\beta>$.
In Eq.~(\ref{b}), $|J,K,M>$  are the familiar rotational states 
of symmetric top, which are characterized by the angular momentum ($J$), 
its projection ($K$) on
the $z$-axis of the molecular system of coordinates and the 
projection ($M$) on the
laboratory quantization axis $Z$. $I$ and $\sigma$ are the total spin of 
the three hydrogen nuclei and its projection on the $Z$-axis,
respectively. The explicit expression for
the spin states $|I,\sigma,K>$ is given in \cite{Townes55}.

Permutation of any two hydrogen nuclei in CH$_3$F inverts 
$z$-axis of the molecular system of coordinates. Consequently, the 
action of such operation  ($P_{23}$, for instance) on the molecular states 
reads: $P_{23}|\beta>=|\overline\beta>$, where  
$\overline\beta\equiv\{J,-K,M,I,\sigma\}$. Note that the complete set of 
the molecular states comprises both $\beta$ and $\overline\beta$
sets.

Using the states $|\beta>$ and $|\overline\beta>$ one can construct the 
states which have the proper symmetry with respect to the permutation of 
any two hydrogen nuclei:
\begin{equation}
              |\beta,\kappa>\ =\ \frac{1}{\sqrt2}
               \left[1 + (-1)^\kappa P_{23}\right]|\beta>;\ \  
               \kappa = 0,1.
\label{bk}
\end{equation}
The action of the permutation of two hydrogen nuclei on the state 
$|\beta,\kappa>$ is defined by the rule:
$P_{23}|\beta,\kappa>=(-1)^\kappa|\beta,\kappa>$ and by similar relations 
for the permutations of the other two pairs of hydrogen nuclei.

In the next step, one has to take into account the symmetric ($|s=1>$)
and antisymmetric ($|s=0>$) inversion states. 
The action of the permutation of the two hydrogen nuclei 
on these states, for example $P_{23}$, reads
\begin{equation}
     P_{23}|s=0>\ =\ -|s=0>;\ \ \ P_{23}|s=1>\ =\ |s=1>.
\label{P}
\end{equation}
Evidently, the cyclic permutation of the three hydrogen nuclei of the 
molecule does not change the inversion states.

The total spin-rotation states of CH$_3$F should be antisymmetric under 
permutation of any two hydrogen nuclei, because protons are fermions.
Consequently, the only allowed states of CH$_3$F are 
$|\beta,\kappa=s>|s>$.

Finally, the description of the CH$_3$F states should be completed 
by adding the spin states of fluorine and
carbon ($^{13}$C) nuclei, both having spin equal 1/2:
\begin{equation}
            |\alpha>\ =\  |\beta,\kappa=s>|s>|\sigma_F>|\sigma_C>,
\label{a}
\end{equation}
where $\sigma_F$ and $\sigma_C$ are the $Z$-projections of the F
and $^{13}$C nuclei' spins, respectively. 
In the following, we will denote the states (\ref{a}) of 
a free molecule as $\alpha$-basis. For the rigid symmetric tops, 
as CH$_3$F is, the states $|\alpha>$ are degenerate over the
quantum numbers $s$, $M$, $\sigma$, $\sigma_F$, $\sigma_C$.

\newpage

Table~1. Experimental and theoretical parameters of the nuclear
spin conversion in $^{13}$CH$_3$F by quantum relaxation.
\vspace{2cm}

\begin{tabular}{|c|c|c|c|c|c|}  \hline
   & $\omega_0/2\pi$ & ${\cal T}$  &  $\Gamma$  & $\gamma(0)$ & $\gamma_b$ \\
   &   (MHz)    &  (kHz)      & (10$^8$~s$^{-1}$/Torr)
                                 &(10$^{-3}$~s$^{-1}$/Torr) 
                                    & (10$^{-3}$~s$^{-1}$/Torr) \\ \hline
Experiment      
   &  132.06 $\pm$ 0.27 & 64.1 $\pm$ 0.5 & 1.9 $\pm$ 0.1 & $12.2\pm0.6^{(4)}$ 
                                         & 4.6 $\pm$ 0.7 \\ \hline
Theory
   &  $130.99\pm0.15^{(1)}$ & $69.2\pm0.2^{(2)}$ & 1.0$^{(3)}$ & 
                                        $12.04\pm0.5^{(5)}$ & -- \\ \hline   
Difference &  1.0 $\pm$ 0.3 & -5.1 $\pm$ 0.5 & -- & 0.15 $\pm$ 0.8 & --\\ 
\hline     
\end{tabular}
\vspace{1cm}

$^{(1)}$Calculated using the molecular parameters from Ref.~\cite{Papousek94JMS},
(Table~1, column~2).

$^{(2)}$Calculated using the molecular structure determined in 
Ref.~\cite{Egawa87JMStr,Egawa98}.

$^{(3)}$The level population decay rate from Ref.~\cite{Jetter73JCP}.

$^{(4)}$Experimental value from Ref.~\cite{Nagels96PRA54}

\begin{figure}[htb]
\centerline{\psfig
{figure=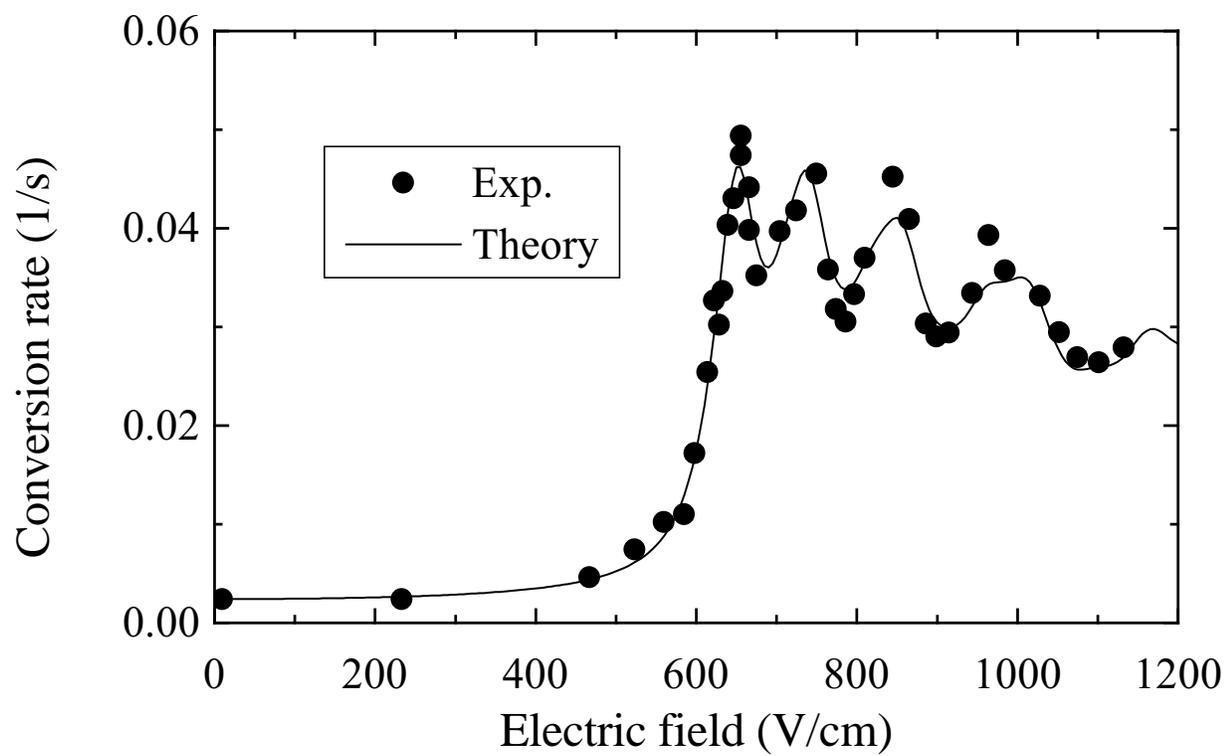,height=12cm}}
\vspace{1cm}
\caption{\sl \label{fig2} Experimental [8] and theoretical 
ortho--para conversion spectrum in $^{13}$CH$_3$F.}
\end{figure}

$^{(5)}$Zero-field value predicted by the theoretical curve in Fig.~\ref{fig2}.

\newpage

\newpage

\end{document}